\documentstyle[12pt,psfig]{article}

\newtheorem{lemma}{Lemma}

\newtheorem{theor}{\large\bf Theorem}

\def\FF{\hbox to 8.33887pt{\rm I\hskip-1.8pt F}}
\def\NN{\hbox to 9.3111pt{\rm I\hskip-1.8pt N}}
\def\PP{\hbox to 8.61664pt{\rm I\hskip-1.8pt P}}
\def\QQ{\rlap {\raise 0.4ex \hbox{$\scriptscriptstyle |$}}
{\hskip -4.5pt Q}}
\def\RR{\hbox to 9.1722pt{\rm I\hskip-1.8pt R}}
\def\ZZ{\hbox to 8.2222pt{\rm Z\hskip-4pt \rm Z}}

\newcommand{\resetequ}{\setcounter{equation}{0}}
\newcommand{\fr}{{\cal F}}           
\newcommand{\tree}{{\cal T}}           

\newcommand{\be}{\begin{equation}}
\newcommand{\ee}{\end{equation}}
\newcommand{\bqa}{\begin{eqnarray}}
\newcommand{\eqa}{\end{eqnarray}}
\newcommand{\ba}{\begin{array}}
\newcommand{\ea}{\end{array}}

\newcommand{\no}{\nonumber}

\newcommand{\lp}{\left (}
\newcommand{\rp}{\right )}
\newcommand{\qed}{\hfill \rule {1ex}{1ex}}

\newcommand{\ep}{\epsilon}

\newcommand{\de}{\delta}

\newcommand{\la}{\lambda}

\newcommand{\si}{\sigma}

\newcommand{\bpsi}{\bar{\psi}}

\begin{document}

\centerline{\large \bf The two dimensional Hubbard Model at half-filling:}
\centerline{\large \bf I. Convergent Contributions}
\vskip 2cm

\centerline{V. Rivasseau}
\centerline{Centre de Physique Th{\'e}orique, CNRS UMR 7644}
\centerline{Ecole Polytechnique}
\centerline{91128 Palaiseau Cedex, FRANCE}

\begin{abstract} We prove analyticity theorems 
in the coupling constant for the Hubbard model at half-filling.
The model in a single renormalization group slice of index $i$ 
is proved to be analytic in $\lambda$ 
for $|\lambda| \le c/i$ for some constant $c$,
and the skeleton part of the model at temperature $T$ 
(the sum of all graphs without two point insertions) is proved to be
analytic in $\lambda$  for $|\lambda| \le c/|\log T|^{2}$.
These theorems are necessary steps towards proving that the Hubbard model at
half-filling is {\it not} a Fermi liquid (in the mathematically precise 
sense of Salmhofer). 
\end{abstract}

\section{Introduction}
\resetequ

Constructive renormalization group approach to the Fermi systems 
of condensed matter [BG][FT1-2] 
is an ongoing program to study quite systematically
the properties of interacting non-relativistic
Fermions at finite density in one, two or three dimensions.
In one dimension interacting Fermions have been proved to form a Luttinger
liquid until zero temperature [BGPS][BM]. 
The simplest interacting two-dimensional model for Fermions, namely the
jellium model, has been recently shown to be a Fermi liquid [DR1-2] 
above the critical temperature, in the
sense of Salmhofer's criterion [S1]. The next most natural model
in two dimensions is the Hubbard model on a square lattice at half-filling
considered in this paper.
This model presents the interesting features of a square Fermi surface
with nesting vectors and van Hove singularities. It has also particle-hole
symmetry, which preserves the Fermi surface under the 
renormalization group flow.
For all these reasons it is the best candidate for a first example in two
dimensions of a Fermionic system which is {\it not} a Fermi liquid,
but rather some kind of Luttinger liquid with logarithmic corrections.

To study the Fermi versus Luttinger behavior according to Salmhofer's 
criterion, one must prove analyticity in the coupling constant
in a domain above some critical temperature, and study whether
the first and second derivatives of the self-energy in Fourier space
are uniformly bounded or not in that analyticity domain\footnote{
In two or more dimensions perturbation theory can generically work
only above some critical temperature, so the Fermi liquid behavior
cannot persist until zero temperature, except for very particular
models with a Fermi surface which is not parity invariant. There is
an ongoing program to study these models in two dimensions [FKLT][FST][S2].}.
This analysis may be conveniently decomposed
into four main steps of increasing difficulty: 

A) control of the model in a single slice

B) control of the model without divergent subgraphs in many slices

C) control of the two point function  renormalization 

D) study of the first and second derivatives of the self-energy

This program is completed only at the moment for the two-dimensional jellium 
model: steps A and B were performed in [FMRT] and steps C and D in [DR2]. 
For the three dimensional jellium model, steps A and B have been
successively completed in [MR] and [DMR]. For the half-filled Hubbard model
we perform in this paper steps A) and B).
We use an angular decomposition of the model
into ``sectors'' that are very different from the jellium case. 
We write the momentum conservation rules in terms of these sectors. 
Then we prove that the sum over all graphs
with momenta restricted to the $i$-th slice of the renormalization
group is analytic for $|\la | \le const/i $ (step A).
We prove two theorems corresponding to step B. The first one states that 
the completely convergent part of the theory,
namely the sum over all graphs which do not contain 
two-particle and four particle subgraphs with external legs closer
to the singularity than their internal legs,
is analytic for $|\la \log T | \le const$. The second result
states that the ``biped-free'' part of the theory,
namely the sum over all graphs which do not contain 
two-particle subgraphs with external legs closer
to the singularity than their internal legs,
is analytic for $|\la \log ^{2}T | \le const$.
We remark that this last domain of analyticity
is the expected optimal domain for the full theory.
We remark also that these domains are smaller
than the domains for the jellium case which are respectively 
$|\la | \le const $ for the single slice or completely convergent
theory, and $|\la \log T | \le const$ for the biped free 
part of the theory or the full theory with an appropriate mass-counterterm.

Finally we remark also that since we expect the half-filled
Hubbard model {\it not} to be a Fermi liquid, step D in that case should
consist of a proof that the second momentum derivative of the self-energy
is {\it  not} uniformly bounded in that domain of analyticity. This requires a
lower bound showing the divergence of this quantity 
near the corner $\la, T \to 0$ rather than an upper bound. 

For a very simple introduction to constructive Fermionic theory,
we recommend [AR1]. We will also use without too much further
explanations the Taylor tree formulas that are
developed in detail in [AR2]. It would also be useful
if the reader has already some familiarity with the
basics of multiscale expansions [R] and
with constructive Fermionic renormalization, as e.g. developed
in [DR3]; but we will try to remain as simple and self-contained as possible. 

\section{Model and Notations}
\resetequ

A finite temperature Fermionic model has a propagator $C(x,\bar{x})$
where $x= (x_{0}, \vec x)$, which is translation invariant.
By some slight abuse of notations we may therefore write it either
$C(x-\bar{x})$ or $C(x,\bar{x})$, where the first point corresponds 
to the field and the second one to the antifield.
This propagator at finite temperature is
antiperiodic in the variable $x_0$ with antiperiod ${1\over T}$, hence its
Fourier transform depends on discrete values (called the Matsubara
frequencies): 
\be 
k_0 = \frac{2n+1}{\beta} \pi \ , \quad n \in \ZZ \ , \label{discretized} 
\ee 
where $\beta=1/T$ (we take $ /\!\!\!{\rm h}
=k =1$). Remark that only odd frequencies appear, because of
antiperiodicity.

The Hubbard model lives on the square lattice $\ZZ^2$, so that 
the three dimensional vector $x = (x_0, \vec{x})$ is such that
$ \vec{x} = (n_1 , n_2 )\in \ZZ^2$.  
From now on we write $v_1$ and $v_2$ for the two components
of a vector $\vec v$ along the two axis of the lattice. 

At half-filling and finite temperature $T$, the Fourier 
transform of the propagator of the Hubbard model is:
\be
\hat{C}_{ab} (k) = \de_{ab} \frac{1}{ik_0-e(\vec{k})},
\quad \quad e(\vec{k})= \cos k_1   + \cos k_2 \ ,
\label{prop}
\ee
where $a,b \in \{\uparrow , \downarrow\}$ are the
spin indices. The vector $\vec k$ lives on the two-dimensional torus 
$\RR^2/ (2 \pi\ZZ)^2$.
Hence the real space propagator is
\be
C_{ab}(x) =\frac{1}{(2\pi)^2\beta}\; \sum_{k_0} \; 
\int_{-\pi}^{\pi} dk_1 \int_{-\pi}^{\pi}dk_2\; e^{ikx}\;
\hat{C}_{ab}(k) \ .
\label{tfprop}\ee

The notation $\sum_{k_0}$ means really the discrete sum over the integer
$n$ in (\ref{discretized}).
When $T \to 0$ (which means $\beta\to \infty$) $k_0$
becomes a continuous variable, the corresponding discrete sum becomes an
integral, and the corresponding propagator $C_{0}(x)$ becomes singular
on the Fermi surface
defined by $k_0=0$ and $e(\vec{k})=0$. This Fermi surface is a square of side
size $\sqrt{2} \pi$
(in the first Brillouin zone) joining the corners $(\pm \pi , 0), (0,\pm\pi )$.
We call this square the Fermi square, its corners and faces are called
the Fermi faces and corners. Considering the periodic boundary conditions,
there are really four Fermi faces, but only two Fermi corners.

In the following to simplify notations we will write:
\be
\int d^3k \; \equiv \; {1\over \beta} \sum_{k_0} \int d^2k
\quad , \quad 
\int d^3x \; \equiv \; {1\over 2}
\int_{-\beta}^{\beta}dx_0 \sum_{\vec{x}\in \ZZ^2} \ . \label{convention}
\ee

In determining the spatial decay we recall that by antiperiodicity
\be
C(x) = f(x_0,\vec{x}) :=
\sum_{m\in \ZZ} (-1)^m \; C_0\lp x_0+{m\over T}, \vec{x}\rp  \ .
\label{copie1}\ee
where $C_0$ is the propagator at $T=0$.
Indeed the function $f$ is 
antiperiodic and its Fourier transform is the right one.

The interaction of the Hubbard model is simply

\be
S_V = \la  \int_V d^3x\; (\sum_a \bpsi\psi)^2 \label{int} \ ,
\ee
where  $V:= [-\beta,\beta]\times V'$ and $ V'$ is an auxiliary finite volume
cutoff in two dimensional space that will be sent later to infinity.
Remark that in (\ref{discretized}) $|k_0|\geq \pi/\beta\neq 0$ 
hence the denominator in $C(k)$ can never be 0 at non zero temperature.
This is why the temperature provides a natural infrared cut-off.

\subsection{Scale Analysis}

The theory has a natural lattice spatial cutoff. 
To implement the renormalization group analysis, we introduce as usually
a compact support function $u(r)\in{\cal C}_{0}^\infty({\rm R})$
(it is convenient to choose it to be Gevrey of order $\alpha<1$ [G]
so as to ensure fractional exponential decrease
in the dual space) 
which satisfies:
\be
u(r)= 0 \quad {\rm for} \ |r|> 2 \ ; \ u(r) =1
\quad {\rm for} \   |r|<1  \ \ . \label{gevrey}
\ee
With this function, given a constant $M\ge 2$, we can construct a partition 
of unity
\bqa 1 &=& \sum_{i=0}^{\infty} u_{i} (r)  \ \ \forall r \ne 0\ \ ; 
\nonumber\\ 
u_0 (r)  &=& 1- u(r) \ ;\  u_{i} (r) \ =\ 
u(M^{2(i-1)}r)-u(M^{2i}r) \ {\rm for}\ 
i\ge  1 \ .
\eqa

The propagator is then divided into slices according to this partition
\be
C(k) = \sum_{i=0}^{\infty} C_i(k)
\ee
where
\be
C_i(k) =  C(k)  u_i [ k_0^2+e^2(\vec{k}) ] \ .
\ee
(indeed $ k_0^2+e^2(\vec{k}) \ge T^{2} > 0$).

In a slice of index $i$ the cutoffs ensure
that the size of $k_0^2 +  e^2(\vec{k})$
is roughly $M^{-2i}$. 
More precisely in the slice $i$ we must have
\be M^{-2i} \le k_0^2 +  e^2(\vec{k})\le 2M^{2}   M^{-2i} \ .
\label{size}
\ee

The corresponding domain is a three dimensional volume whose section
through the $k_0=0$ plane is the shaded region pictured in Figure 1.

\begin{figure}
\centerline{\psfig{figure=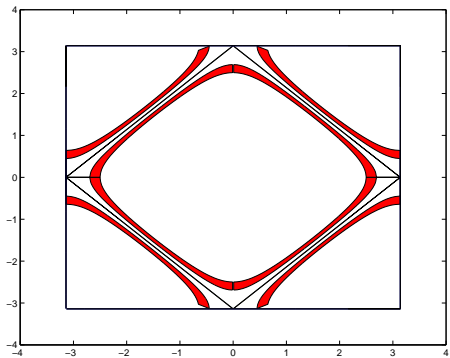,width=6cm}}
\caption{A single slice of the renormalization group}
\label{oslic}
\end{figure}

Remark that at finite temperature, the propagator $C_i$ vanishes
for $i\ge i_{max}(T)$ where $M^{i_{max}(T)}\simeq 1/T$ (more
precisely  $i_{max}(T) = E( \log {M \sqrt 2 \over \pi T }/\log M)$, 
where $E$ is the integer part), so there is only a finite number of 
steps in the renormalization group analysis. 

Let us state first our simplest result,
for a theory whose propagator is only $C_i$, hence corresponds to a generic
step of the renormalization group\footnote{In the following we assume $i\ge 1$.
Indeed the first slice $i=0$ is somewhat peculiar because of
the unboundedness of the Matsubara frequencies, which requires
a little additional care.}: 

\begin{theor} The Schwinger functions of the theory with propagator
$C_i$ and interaction (\ref{int}) are analytic in $\lambda$ in a disk
of radius $R_i$ which is at least $c/i$ for a suitable constant $c$:
\be  R_i  \ge c /i \ . 
\ee 
\label{radoneslice}
\end{theor}

The rest of this section is devoted to definitions and preliminary lemmas
about sectors, their scaled decay and momentum conservation rules.
Although Theorem 1 applies to a single slice, its proof nevertheless requires
some kind of multiscale analysis, which is done in Section III. 
Our next results, Theorem 2 and 3, which bound the sum
over all ``convergent contributions'', that is without divergent
two point insertions, are slightly more technical to state, but their proof
is almost identical to that of Theorem 1. They are postponed to section IV.

As discussed in the introduction this result is a first step towards
the full analysis of the model in the regime 
$|\lambda \log^2 T | \le const.$, and a rigorous proof that it is not
a Fermi liquid in the sense of Salmhofer.

\subsection{Sectors}

The "angular" analysis is completely different from the jellium
case. We remark first that in our slice, $k_0^2 + e^2(\vec{k})$
is of order $M^{-2i}$, but this does not fix the size of
$e^2(\vec{k})$ itself, which can be of order $M^{-2j}$
for some $j \ge i$. In order for sectors defined in momentum space to 
correspond to propagators with dual decay in direct space,
it is essential that their length in the tangential direction is not too big,
otherwise the curvature is too strong
for stationary phase methods to apply. This was discussed first in [FMRT].
This leads us 
to study the curve $(\cos k_1 + \cos k_2)^{2} = M^{-2j} $ for arbitrary
$j\ge i$. We can by symmetry restrict ourselves to the region 
$0\le k_1 \le \pi/2$, $k_2 >0$.
It is then easy to compute the curvature radius of that curve, which is

\be  R = {(\sin^2 k_1 + \sin^2 k_2)^{3/2} \over 
| \cos k_1 \sin^2 k_2 + \cos k_2 \sin^2 k_1 |}\ .
\ee
We can also compute the distance $d(k_1) $ 
to the critical curve $ \cos k_1 + \cos k_2 =0$,
and the width $w(k_1)$ of the band 
$M^{-j} \le  |\cos k_1 + \cos k_2 |\le \sqrt{2}M. M^{-j} $.
We can then easily check that 
\be d(k_1) \simeq w(k_1) \simeq {M^{-j} \over M^{-j/2} + k_1 } \ ,
\ee
\be  R(k_1) \simeq   {k_1^3 + M^{-3j/2} \over M^{-j}} \ ,
\ee
where $f\simeq g$ means that on the range $0\le k_1 \le \pi/2$ we have
inequalities $ cf \le g\le df$ for some constants $c$ and $d$.

Defining the anisotropic length 
\be  l(k_1) = \sqrt{ w(k_1)R(k_1) } \simeq M^{-j/2} + k_1 \ ,
\ee
the condition in [FMRT] is that the sector length should
not be bigger than that anisotropic length. This leads to the idea that
$k_1$ or an equivalent quantity should be sliced according to a geometric
progression from 1 to $M^{-j/2}$ to form the angular sectors in this model.

For symmetry reasons
it is convenient to introduce a new orthogonal but not normal 
basis in momentum space $(e_+, e_-)$,
defined by $e_+ = (1/2)(\pi, \pi)$ and $e_- = (1/2)(-\pi, \pi)$.
Indeed if we call $(k_+, k_{-})$ the coordinates
of a momentum $k$ in this basis, the Fermi surface is given by
the simple equations $k_+ = \pm 1$ or  $k_- = \pm 1$.
This immediately follows from the identity
\be \cos k_1 + \cos k_2  = 2 \cos (\pi  k_{+}/2 )
\cos (\pi  k_{-}/2 ) \ .
\ee
(Note however that the periodic b.c. are more complicated in that new basis).
Instead of slicing $e(\vec k ) $ and $k_1$, it is then more symmetric to
slice directly $\cos (\pi  k_{+}/2 )$ and $\cos (\pi  k_{-}/2 )$.

Guided by these considerations we introduce the partition of unity
\be
1 = \sum_{s=0}^{i} v_{s}(r) \ ; 
\cases{v_{0}(r) = 1- u(M^{2}r) \cr
v_{s}= u_{s+1}\ & for  $1\le s\le i-1$ \cr
v_{i}(r) = u(M^{2i}r)\cr  }
\ee
and define
\be
C_{i}(k) = \sum_{\sigma = (s_{+}, s_{-})} C_{i,\sigma}(k)
\ee
where
\be
C_{i,\sigma}(k)=  C_i(k) v_{s_{+}} [ \cos^{2} (\pi k_{+} /2 ) ]\;
v_{s_{-}} [ \cos^{2}\pi k_{-}/2) ] \ .
\ee
We remark that using (\ref{size}) in order for $C_{i,\sigma}$ not to be 0, 
we need to have $s_{+} + s_{-} \ge i-2$. We define the ``depth''
$l(\sigma)$ of a sector to be $l = s_{+} + s_{-} - i + 2$.

To get a better intuitive picture of the sectors,
we remark that they can be classified into different categories:

- the sectors (0,i) and (i,0) are called the middle-face sectors

- the sectors (s,i) and (i,s) with $0 <s<i$ are called the face sectors

- the sector (i,i) is called the corner sector

- the sectors (s,s) with $ (i-2)/2 \le s <i$ are called the diagonal sectors

- the others are the general sectors

Finally the general or diagonal
sectors of depth 0 for which $s_{+} + s_{-} = i-2$ 
are called border sectors.

If we consider the projection onto the $(k_{+}, k_{-})$ plane,
taking into account the periodic b.c. of the Brillouin zone,
the general and diagonal sectors have 8 connected components, the 
face sectors have 4 connected components, the middle
face sectors and the corner sector have 2 connected components. 
In the three dimensional space-time, 
if we neglect the discretization of
the Matsubara frequencies, these numbers would double 
except for the border sectors.

\subsection{Scaled decay}

\begin{lemma} Using Gevrey cutoffs of degree $\alpha <1$, 
the propagator $C_{i,\si}$ obeys the scaled decay
\be | C_{i,\si} | \le  c. M^{-i-l} e^{-[d_{i,\si } (x,y)]^{\alpha}}
\label{decay}
\ee
where
\be  d_{i,\si } (x,y) = \{ M^{-i}|x_0 -y_0| +  
M^{-s_{+} }|x_{+} -y_{+}| +
M^{-s_{-}}|x_{-} -y_{-}| \}  \ .
\ee
\end{lemma}

\noindent{\it Proof} This is essentially Fourier analysis 
and integration by parts.
If $x = (n_1, n_2)\in \ZZ^{2}$, we define $(x_{+}, x_{-})= (\pi/2)(n_1+n_2,
n_2 -n_1)$. The vector $(x_{+}, x_{-})$ then belongs to $(\pi/2)\ZZ^{2}$
but with the additional condition that $x_{+}$ and $x_{-}$ have the same 
parity. 

Defining, for $X \in [(\pi/2)\ZZ\;]^{2}$
\bqa  D_{i,\sigma}( X) = && (1/2) {1 \over 8 \beta} \sum_{k_0}
\int_{-2}^{+2} dk_{+}\int_{-2}^{+2} dk_{-}
e^{i (k_{0}x_{0}+ k_{+}x_{+} + k_{-}x_{-})}  
\nonumber \\
&& {u_i [ k_0^2+  4\cos^{2}(\pi k_{+}/2) \cos^{2} (\pi k_{-}/2) ] 
\over ik_{0} - 2\cos(\pi k_{+}/2) \cos (\pi k_{-}/2) } 
\nonumber \\
&& v_{s_{+}} [ \cos^{2} (\pi k_{+} /2 ) ]\;
v_{s_{-}} [ \cos^{2}(\pi k_{-}/2) ]
\label{doubl}
\eqa
we note that $C_{i,\sigma}(X)= D_{i,\sigma}( X)$ for $X$ satisfying
the parity condition.

(Remember the Jacobian ${\pi ^{2} \over 2}$ from $dk_1 dk_2$ to 
$dk_+ dk_-$, and the initial domain of integration that is doubled.)

The volume of integration trivially gives a factor $M^{-i}$ for the 
$k_{0}$ sum and factors $M^{-s_{+}}$ and $M^{-s_{-}}$ for the
$k_{+}$ and $k_{-}$ integration (see (\ref{supp}) below). The integrand
is trivially bounded by $M^{i}$ on the integration domain, and this explains
the prefactor $cM^{-i-l}$ in (\ref{decay}).

We then apply standard integration by parts techniques to formulate
the decay. From e.g. Lemma 10 in [DR1] we know that to obtain
the scaled decay of Lemma 1 we have only to
check the usual derivative bounds in Fourier space:

\bqa \Vert {\partial^{n_0} \over \partial k_0^{n_0} }
{\partial^{n_+} \over \partial k_+^{n_+} }
{\partial^{n_-} \over \partial k_0^{n_-} } \hat D_{i,\sigma} \Vert
\le A.B^{n} M^{i n_0} M^{s_{+}n_+} M^{s_{-}n_-} 
(n !)^{1/\alpha} \label{foubou}
\eqa
where $n=n_{0}+n_{+}+n_{-}$, and the derivative 
${\partial \over \partial k_0}$ really means the 
natural finite difference operator $(1/2\pi T) (f(k_0 +2\pi T)-f(k_0))$ 
acting on the discrete Matsubara frequencies. The norm is the ordinary sup
norm.
 
But from (\ref{doubl}), 
\bqa\hat D_{i,\sigma} (k)&=& {1 \over 16\beta}  
{u_i [ k_0^2+  4\cos^{2}(\pi k_{+}/2) \cos^{2} (\pi k_{-}/2) ] 
\over ik_{0} - 2\cos(\pi k_{+}/2) \cos (\pi k_{-}/2) } 
\nonumber\\
&&v_{s_{+}} [ \cos^{2} (\pi k_{+} /2 ) ]\;
v_{s_{-}} [ \cos^{2}(\pi k_{-}/2) ]
\eqa
and the derivatives are bounded easily using the standard
rules for derivation, product and composition of Gevrey functions,
or by hand, using the support properties of the 
$v_{s_{+}}$ and $v_{s_{-}}$ fonctions. For instance a derivative
${\partial \over \partial k_+}$ can act on the 
$v_{s_{+}} [ \cos^{2} (\pi k_{+} /2 ) ]$ factor, in which case it 
is easily directly bounded by $c M^{s_{+}}$ for some constant $c$. 
When it acts on
$u_i [ k_0^2+  4\cos^{2}(\pi k_{+}/2) \cos^{2} (\pi k_{-}/2) ] $
it is easily bounded by  $c.M^{2i-s_+-2s_-}$ 
hence by $c. M^{s_{+}}$, using the relation $s_{+}+ s_{-}\ge i-2$.
When it acts on the denominator
$[ik_{0} - 2\cos(\pi k_{+}/2) \cos (\pi k_{-}/2) ]^{-1} $.
it is bounded by $c.M^{i-s_-}$, hence again 
by $c. M^{s_{+}}$, using the relation $s_{+}+ s_{-}\ge i-2$.
Finally when it acts on a  $\cos (\pi k_{+} /2 ) $ created
by previous derivations, it costs directly $c. M^{s_{+}}$.
The factorial factor $(n !)^{1/\alpha}$ in (\ref{foubou})
comes naturally from deriving the cutoffs,
which are Gevrey functions of order $\alpha$; deriving other factors
give smaller factorials (with power 1 instead of $1/\alpha$).

Finally a last remark: to obtain the Lemma for the last slice, 
$i=i_{max}(T)$, one has to take into account the fact 
that $x_{0}$ lies in a compact circle, so that there is really no 
long-distance decay to prove.

\subsection{Support Properties}

If $C_{i,\si } (k) \ne 0$, the momentum $k$ must obey
the following bounds:

\be |k_0| \le \sqrt 2 M  M ^{-i}
\ee
\be \cases{ M^{-1}  \le  |\cos (\pi k_{\pm} /2)| \le 1
\ & for  $s_{\pm} = 0 $  \ ,\cr
M^{-s_{\pm}-1}  \le  |\cos (\pi k_\pm /2)| \le \sqrt 2 M ^{-s_{\pm}}
\ & for  $1\le s_{\pm}\le i-1$  \ ,\cr
|\cos (\pi k_\pm /2)| \le \sqrt 2 M ^{-i}
\ & for  $s_{\pm} = i$  \ .\cr }
\ee
In the support of our slice in the first
Brillouin zone we have $ |k_+ | < 2$ and 
$|k_-| < 2$ (this is not essential but the inequalities 
are strict because $i\ge1$). 
It is convenient to associate to any such component
$k_{\pm}$ a kind of ``fractional part'' called $q_{\pm}$ defined by
$q_{\pm}= k_{\pm}-1$ if $k_{\pm}\ge 0$ and  
$q_{\pm}= k_{\pm}+1$ if $k_{\pm}<0$, 
so that $0\le |q_{\pm}| \le 1 $. 
Then the bounds translate into
\be \cases{  2/ \pi M   \le | q_\pm  | \le 1
\ & for  $s_{\pm} = 0 $  \ ,\cr
2 M^{-s_{\pm}} / \pi M  \le | q_\pm  | \le \sqrt 2
M ^{-s_{\pm }} 
\ & for  $1\le s_{\pm }\le i-1$  \ ,\cr
| q_\pm   |\le \sqrt 2  M ^{-i} 
\ & for  $s_{\pm } = i$  \ .\cr }
\label{supp}
\ee

\subsection{Momentum conservation rules at a vertex}

Let us consider that the four momenta $k_1$, $k_2$, $k_3$, $k_4$,
arriving at a given vertex $v$
belong to the support of the four sectors 
$\si_1$, $\si_2$, $\si_3$, $\si_4$, in slices $i_1$, $i_2$, $i_3$, $i_4$.
In Fourier space the vertex (\ref{int}) implies constraints
on the momenta. Each spatial component of the sum of the four momenta 
must be an integer multiple of $2\pi$ in the initial basis, 
and the sum of the four Matsubara frequencies must also be zero.
 
In our tilted basis $(e_{+},e_{-})$, this translates into the conditions:

\be k_{1,0} + k_{2,0} + k_{3,0} + k_{4,0} = 0 \ ,
\ee
\be k_{1,+} + k_{2,+} + k_{3,+} + k_{4,+} = 2n_{+}  \ ,\label{consplus}
\ee
\be k_{1,-} + k_{2,-} + k_{3,-} + k_{4,-} = 2n_{-}  \ ,\label{consminus}
\ee
where $n_{+}$ and $n_{-}$ must have identical parity.

We want to rewrite the two last equations in terms of the 
fractional parts $q_{1}$, $q_{2}$, $q_{3}$ and $q_{4}$. 

Since an even sum of integers which are $\pm 1$ is even, we find
that (\ref{consplus}) and (\ref{consminus}) imply  
\be q_{1,+} + q_{2,+} + q_{3,+} + q_{4,+} = 2m_{+}  \ ,\label{coplus}
\ee
\be q_{1,-} + q_{2,-} + q_{3,-} + q_{4,-} = 2m_{-}  \ ,\label{cominus}
\ee
with $m_{+}$ and $m_{-}$ integers. Let us prove now that except
in very special cases, these integers must be 0. Since $|q_{j,\pm}|\le 1$,
$|m_{\pm}|\le 2$. But  $|q_{j,\pm}| = 1$ is possible only for $s_{j,\pm}=0$.
Therefore $|m_{\pm}| = 2$ implies $s_{j,\pm}=0 \ \forall j$.
Now suppose e.g. $|m_{+}|=1$. 
Then  $s_{j,+}$ is 0 for at least two values of 
$j$. Indeed for $s_{j,\pm}\ne 0$ we have $|q_{j,\pm}|\le \sqrt 2 M^{-1}$,
and assuming $3 \sqrt 2 M^{-1} < 1$, equation (\ref{coplus}) could not hold.

We have therefore proved

\begin{lemma}
$m_{+}=0$ unless $s_{j,+}$ is 0 for at least two values of 
$j$, and $m_{-}=0$ unless $s_{j,-}$ is 0 for at least two values of 
$j$.
\end{lemma}

Let us analyze in more detail equations (\ref{coplus}) and 
(\ref{cominus}) for $|m_{+}|=|m_{-}|=0$. Consider e.g.
(\ref{coplus}). By a relabeling
we can assume without loss of generality that
$s_{1,+} \le s_{2,+} \le s_{3,+}\le s_{4,+}$
Then either $s_{1,+}= i_1 $ or $s_{1,+} < i_1$, in which case
combining equations (\ref{coplus}) and (\ref{supp}) we
must have:
\be 3\sqrt 2 M^{-s_{2,+}} \ge 2 M^{-s_{1,+}} /\pi M \ ,
\ee
which means
\be s_{2,+} \le s_{1,+} + 1 + {\log (3\pi/\sqrt 2) \over \log M} \ .
\ee
This implies
\be | s_{2,+} - s_{1,+}| \le 1 
\ee
if $M > 3 \pi /\sqrt 2 $, which we assume from now on.

The conclusion is:

\begin{lemma}
If $m_{\pm}=0$, either the smallest index $s_{1,\pm}$ 
coincides with its scale $i_1 $, or the two smallest
indices among $s_{j,\pm}$ differ by at most one unit.
\end{lemma}

Now we can summarize the content of both Lemmas in a slightly weaker 
but simpler lemma:

\begin{lemma}
\noindent

\noindent{\bf A) (single slice case)}

\medskip
The two smallest
indices among $s_{j,+}$ for $j=1,2,3,4$ differ by at most one unit,
and the two smallest
indices among $s_{j,-}$ for $j=1,2,3,4$ differ by at most one unit.

\medskip
\noindent
{\bf B) Multislice case}

The two smallest
indices among $s_{j,+}$ for $j=1,2,3,4$ differ by at most one unit
or the smallest one, say $s_{1, +}$ must coincide with its scale $i_1$, 
which must then be strictly smaller than the three other scales $i_2$, $i_{3}$
and $i_{4}$. Exactly the same statement holds independently 
for the minus direction.
\label{consmom}
\end{lemma}

\section{The expansion}

For simplicity let us prove the theorem for the pressure:
\be p = \lim_{V\to \infty} {1\over |V|} \log Z (V)\ ,
\ee 
\be Z (V) = \int d\mu_{C_{i}} (\bpsi, \psi) 
e^{\la\int_V d^3x\; (\sum_a \bpsi\psi)^2 (x)}
\ee
where $d\mu_{C_{i}} (\bpsi, \psi) $ is the Grassmann Gaussian measure
of covariance $C_{i}$. (The proof extends without difficulty to any 
Schwinger function at fixed external momenta).

We develop each field and antifield into a sum
over sectors, obtaining a collection of sectors $\{\si \}$.
For each vertex $j$ there are four field or antifields, hence four sectors
called $\si^1_{j}$, $\si^2_{j}$, $\si^3_{j}$ and $\si^4_{j}$.
Integrating over the Grassmann measure, $Z(V)$ becomes:
\bqa
&&Z(V) = \sum_{n} {\la^n \over n!}  
\int_{V^{n}} d^3x_{1}...d^3x_{n} \sum_{a_{j},b_{j}}\sum_{\{\si \}}
\no\\
&&
\left \{\ba{ccccc}
x_{1, a_1, \si^1_{1}}&
x_{1,b_1, \si^2_{1}}&...&x_{n,a_n, \si^1_{n}}&x_{n,b_n,\si^2_{n}}\\
x_{1, a_1, \si^3_{1}}&
x_{1,b_1,\si^4_{1}}&...&x_{n,a_n,\si^3_{n}}&x_{n,b_n,\si^4_{n}}\\
\ea
\right \}\no
\eqa
where we used Cayley's notation for the determinants:
\be
\left \{\ba{c}
u_{j,a,\si}\\ v_{k,b,\si '}\\ \ea \right \} = 
\det (\de_{ab}\de_{\si \si'} C_{i,\si} (u_j - v_k))
\label{det}
\ee
and $a_j,b_j$ are the spin indices.

We know that if we expand the Cayley determinant
the pressure is given by the sum over all connected vacuum graphs 
with one particular vertex fixed at the origin (using translation
invariance). But this formula is not suited for convergence. Instead 
we want to connect the vertices of the connected vacuum graphs
by a tree formula, because these formulas together
with Gram's inequality on the remaining determinant 
are the most convenient to prove convergence [L][AR1]. 
However we want this formula
ordered with respect to increasing
values of the depth index $l= s_{+} + s_{-} -i +2$
(which runs between 0 and $i+2$), so that tree lines with lowest
depth are expanded first. This is conveniently
done using the Taylor jungle formula [AR2, Theorem IV.3].
We obtain:

\bqa 
p &=& \sum_{n} p_{n} \la^n \\
p_{n} &=& { 1\over n!} \sum_{\{a,b, \si\}}
\sum_{{\cal J}} \ep ({\cal J}) \prod_{j=1}^{n} \int dx_{j} \de(x_1) 
\no \\
&&\prod_{\ell\in \tree} \int_{0}^{1} dw_{\ell}
C_{i,a_{\ell},\si_{\ell}} (x_{\ell}, \bar x_{\ell}) 
\det\nolimits_{{\rm left}} (C_{i,\si}(w)) \label{detleft} 
\eqa
where ${\cal J} = (\fr_0 \subset \fr_1 \subset ... \subset  \fr_{i+1}
\subset\fr_{i+2}=\tree )$
is a layered object called a jungle in [AR2]),
$\ep ({\cal J}) $ being an inessential sign. Such a jungle is an increasing 
sequence of forests $\fr_{j}$. A forest is simply a set of lines
which do not make loops, hence in contrast with a tree
it can possibly have several connected components. 
Here the sum is constrained over the jungles whose
last layer $\fr_{i+2}=\tree $ must be a real tree $\tree$ connecting 
the $n$ vertices. This constraint arises because we are computing a connected
function, namely the pressure.
The notation $\det_{{\rm left}} (C_{i}(w)) $ means the determinant 
made of the fields and antifields left after
extraction of the tree propagators. It is therefore a $n+1$ by
$n+1$ square matrix of the Cayley type similar to (\ref{det}),
but with an additional multiplicative parameter depending on the interpolating
parameters $\{w\}$. More precisely, its $(f,g)$ entry 
between field $f$ and antifield $g$ is zero unless
the spin and sectors for $f$ and $g$ coincide. 
In that case let $\chi (f,j)$ be 1 if field $f$ hooks to vertex $j$ and 
zero otherwise. Let also $l_f$ be the depth of
a field or antifield $f$.
The $(f,g)$ entry of the determinant left in  (\ref{detleft})
is then:
\bqa
C_{i}(w)_{fg} &=&  \de_{\si (f)\si(g)}\de_{a(f)b(g)} 
\sum_{j=1}^{n}\sum_{k=1}^{n}
\chi (f,j)\chi (g,k) 
\no\\
&& w^{{\cal J},l_f}(j,k)(\{w\}) 
C_{i,a(f),\si (f)} (x_j,x_k)
\eqa
where
$w^{{\cal J},l}(j,k)(\{w\})$ is given by a rather complicated formula:

- If the vertices $j$ and $k$
are not connected by $\fr_{l}$, then $w^{{\cal J},l}(j,k)(\{w\})= 0$

- If the vertices $j$ and $k$
are connected by $\fr_{l-1}$, then $w^{{\cal J},l}(j,k)(\{w\})= 1$

- If the vertices $j$ and $k$
are connected by $\fr_{l}$, but not by $\fr_{l-1}$, 
then $w^{{\cal J},l}(j,k)(\{w\})$ is the infimum of the $w_{\ell}$ parameters
for $\ell$ in the unique path in the reduced forest 
$\fr_{l}/\fr_{l-1}$ 
connecting the two vertices [AR2]\footnote{The reduced
forest $\fr_{l}/\fr_{l-1}$ is as usual 
the one in which all the connected components of $\fr_{l-1}$
have been contracted to a single vertex.}. The natural convention is that
$\fr_{-1}=\emptyset$ and that $w^{{\cal J},l}(j,j)=1$.

We will only need to know that the matrix $w^{{\cal J},l}(j,k)(\{w\})$  
is a positive $n$ by $n$ matrix with entries labeled by 
the {\it vertices} $j$ and $k$.
This is enough to bound these interpolation parameters by 1
in Gram's bound for the $\det\nolimits_{{\rm left}} (C_{i}(w))$.
This is explained in detail in [AR1] and [DR3].

Now at any given level $l$ the forest $\fr_{l}$ defines a certain set
of $c(l)$ different connected components
(some of them eventually reduced to a trivial isolated vertex). 
To each such connected component
correspond a subgraph, called $G^k_{l}$, $k=1,2,..., c(l)$, which has
a well defined number of internal vertices and a well defined
even number of external fields $e(G_l^k)$. These external fields
are the fields of index greater than $l$ hooked to the internal vertices,
which are themselves joined together
by the forest $\fr_{l}$. These connected components $G^k_{l}$ play a
fundamental role in any multislice analysis [R]; their inclusion
relations form an other tree, the so-called Gallavotti-Nicol{\`o} tree.   

Remark that the
final tree $\tree$ plus the collection $\{\si \}$ of sectors for all
fields obviously determine the full layered tree structure
${\cal J}$ and the connected components $G^k_{l}$
at level $l$, together with their number of external legs 
$e(G_l^k)$. Hence the sums over $a,b, \si$ and ${\cal J}$ in (\ref{detleft})
are redundant, and can be replaced by a simpler sum 
over $a,b, \si$ and $\tree$.

Anticipating on what follows, the power counting of the two 
point connected components $G_l^k$ (those for which $e(G_l^k)=2$)
is marginal. We will need to identify the pair of
external fields of these components
also called ``bipeds'', in order to take into account their extra momentum 
conservation rule.
The bipeds $b$ (together with the full final graph
which we call $G$) form a tree for the inclusion relation, called
${\cal B}$. This tree is a subtree of the ``clustering
tree structure'' [DR1] or ``Gallavotti-Nicolo'' tree, whose nodes are 
the connected components $G^k_{l}$ and whose lines depict 
their inclusion relation (the Gallavotti-Nicolo tree is not to 
be confused with $\tree$).

\begin{figure}
\centerline{\psfig{figure=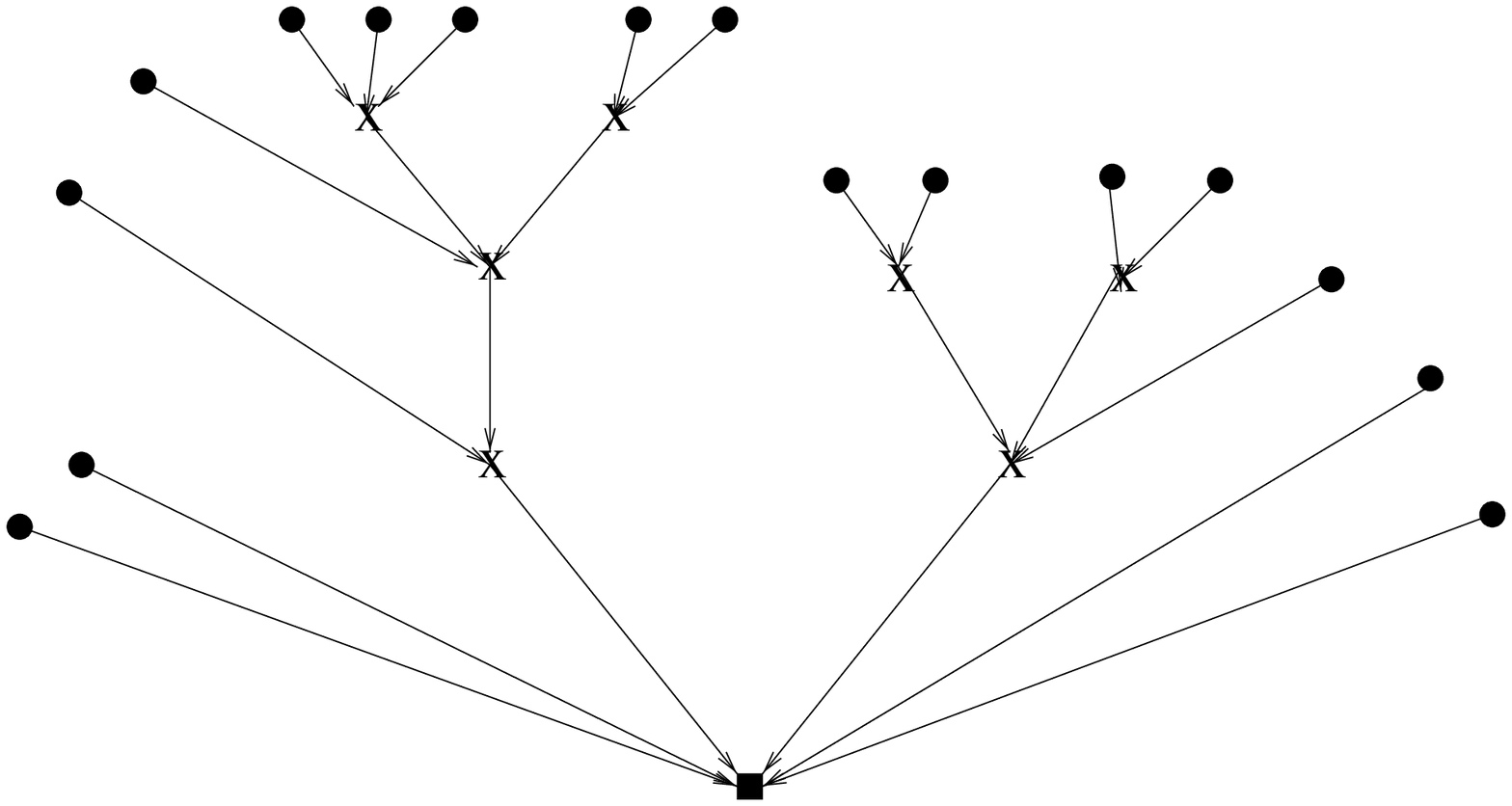,width=12cm}}
\caption{The forest $ {\cal B}$}
\label{bipf}
\end{figure}

We can picture the tree ${\cal B}$ as follows:
every biped is pictured as a cross, every bare vertex as a dot. There
is an inclusion line from each dot to the smallest biped containing
it, and from each biped to the unique next bigger biped containing it.
These inclusion lines which form the forest ${\cal B}$ are pictured as 
downwards arrows in Figure \ref{bipf}. To recover finally a tree,
the last vertex or root, pictured as a box, corresponds to the full graph
$G$ which contains all the maximal bipeds and the remaining dots
(in our case of the pressure, it  cannot be a biped itself 
since it is a vacuum graph). 

For each biped $b\in {\cal B}$ we also fix the two external fields
$\bar\psi_b$ and $\psi_b$ of the biped. 
They must be hooked to two vertices in $b$, $v_{b}$ and $\bar v_{b}$
hence to two dots for which the path to the root using the downwards arrows
in Figure \ref{bipf} passes through $b$. 
Remark indeed that $\bar\psi_b$ and $\psi_b$ must be hooked
to two different vertices, since tadpoles obviously vanish in this theory
at half-filling (by the particle-hole symmetry). Obviously also
the sectors for the fields $\bar\psi_b$  and $\psi_b$,
namely $\bar\si_b$ and $\si_b$ must have the largest depth index $l$
among the four sectors hooked respectively to  $v_{b}$ and $\bar v_{b}$,
otherwise $b$ would not be a connected component $G^{k}_{l}$ for some $l$.
By exact momentum conservation, the external momentum
of the biped must belong to the support of these two sectors. 
Hence they must have equal or
neighboring indices $s_{+}$ and $s_{-}$. 
Also when $b$ varies,
the fields $\bar\psi_b$  and $\psi_b$, and also the
vertices $v_{b}$ and $\bar v_{b}$ are all disjoint. This is a more
subtle property. It is true because by momentum conservation, 
a field cannot be an external field for two bipeds at two different scales
(since, necessarily, the biggest would be one-particle reducible, and
momentum conservation would be violated)\footnote{Strictly speaking,
since our $C^{\infty}_{0}$ 
cutoffs have some overlap, this is true only if we define
a biped as a component $G^{k}_{l}$ with the external scale
at least equal to the maximal internal scale plus 2 (not plus 1),
that is with a strict gap between internal and external scales. 
This inessential complication is left to the reader.
The two point connected components
without such a strict gap do not create any divergence at all.
They can be treated therefore as ordinary connected components,
with more than two external legs, in the power counting below.}. 
 
The set of these data $(\bar\psi_b, \psi_b)$
for all $b\in {\cal B}-\{G\}$ is denoted ${\cal E B}$. By the previous
remarks, it can be described as an even set $V$ (the external vertices
of the bipeds), plus a partition of this set into pairs
$v_{b}$ and $\bar v_{b}$, one for
each $b$, and, again for each $b$, the choice of one field $\psi_{b}$
hooked to $v_b$ and one antifield $\bar\psi_{b}$
hooked to $\bar v_b$.

We now fix ${\cal B}$, ${\cal E B}$, $\{a,b\}$, and $T$, and sum over 
those $\{ \si\}$ that give rise to these data. The constraint 
that  $\{ \si\}$ give rise to these data is indicated
by a prime in the corresponding sum. Remark in particular the
constraint that for any $b\in {\cal B}$ the depth $l(\psi_{b})$ 
must be maximal among the four depths $l_{1},l_{2},l_{3},l_{4}$
of the sectors hooked to $v_{b}$ and the depth $l(\bar\psi_{b})$ 
must be maximal among the four depths $l_{1},l_{2},l_{3},l_{4}$
of the sectors hooked to $\bar v_{b}$, otherwise the subgraph $b$ would
not appear as a connected component $G^{k}_{l}$.
  
To take into account the momentum conservation constraints we introduce 
now for each vertex the function 
$\chi_{j}(\si)= \chi (\si^1_j, \si^2_j, \si^3_j, \si^4_j)$ which is
1 if the condition of Lemma 4 is satisfied and 0 otherwise. We introduce
also for each two-point subgraph $b$ of the forest ${\cal B}$
the constraint $\chi_{b}(\si) $ 
that states that the sectors of its two external legs $\si_b$
and $\bar\si_b$ must overlap, that is must be equal or nearest neighbors. 
These insertions are free since the contributions
for which these $\chi $ functions are not 1 are zero.
They must be done before taking the Gram bound, which destroys the Fourier 
oscillations responsible for momentum conservation at each vertex. We get: 

\bqa 
p_n &=& {1 \over n!}\sum_{{\cal B}, {\cal E B} \atop \{a,b\}, \tree} \ 
\sum'_{\{\si\}}
\ep (\tree) \prod_{j=1}^{n}   \int dx_{j}  \de(x_1)
\prod_{\ell\in \tree} \int_{0}^{1} dw_{\ell}
C_{i,a_{\ell},\si_{\ell}} 
(x_{\ell}, \bar x_{\ell}) \no\\
&&\prod_{j=1}^{n} 
\chi_{j}(\si) \prod_{b \in {\cal B}} \chi_{b}(\si)   
\det\nolimits_{{\rm left}} (C_{i}(w)) \ . \label{form} 
\eqa

We apply now Gram's inequality to the determinant as explained in detail 
in [DR3]. For that purpose we rewrite $C_{i,\si}$ as a product of
two half propagators 
in Fourier space. Taking the square root of the positive matrix
$w$ we obtain the bound:
\be
\det\nolimits_{{\rm left}} (C_{i}(w)) \le c^{n} \prod_{f \ {\rm left}}
M^{-(i+l_{f})/2}
\ee
where the product runs over all fields and antifields left by the tree
expansion. Indeed the half-propagators corresponding
to $C_{i,\si}$ may be chosen to contribute each to one half
of the full propagator scaling factor  
$M^{-i-l_f}$ in (\ref{decay}).

We can now integrate over the positions of the vertices save the fixed one
$x_1$ using the Gevrey scaled decay (\ref{decay})
and obtain a bound on the $n-th$ order of perturbation theory

\be | p_n | \le {c^n \over n!} M^{-2i}
\sum_{\bar {\cal B}, {\cal E B}\atop \{a,b\}, \tree}\ \sum'_{\{\si \}} 
\prod_{j=1}^{n} \chi_{j}(\si) \prod_{b \in {\cal B}} \chi_{b}(\si)  
\prod_{\ell \in \tree} M^{l_{\ell}}
\prod_{f} M^{-l_{f}/2} \label{absol1}
\ee
where the product over $f$ now runs over all the $4n$ fields
and antifields of the theory.

We can check by induction that:
\be \prod_{f} M^{-l_{f}/2} = \prod_{l=0}^{i+2}\prod_{k} M^{-e(G_l^k)/2}\ ,
\label{induc1}
\ee
\be \prod_{\ell \in \tree} M^{l_{\ell}} =  M^{-i-3}
\prod_{l=0}^{i+2}\prod_{k}  M^{1}\ .
\label{induc2}
\ee
(to prove the last equality, remember that $\tree$ is a subtree in
each connected component $G_l^k$).
We obtain the bound
\be | p_n | \le {c^n \over n!} M^{-3i-3}
\sum_{\bar {\cal B}, {\cal E B}\atop \{a,b\}, \tree}\ \sum'_{\{\si \}} 
\prod_{j=1}^{n} \chi_{j}(\si)  
\prod_{b \in {\cal B}} \chi_{b}(\si) 
\prod_{l=0}^{i+2}\prod_{k}  M^{1-e(G_l^k)/2}\ .
\ee

Therefore we have exponential decay 
in index space except for the bipeds $b\in {\cal B}$.
Indeed for $e\ge 4$ we have $e/2 -1 > e/4$.
At each vertex $j$ we have four sectors with depths $l_j^1$, $l_j^2$, $l_j^3$ 
and $l_j^4$. The data in ${\cal EB}$ in particular contain
the information about the set $V$ of vertices for which 
the line with maximal index,
$l_j^4$, is the external line of a biped in ${\cal B}$. Therefore we obtain:

\be | p_n | \le{c^n \over n!} M^{-3i-3}
\sum_{\bar {\cal B}, {\cal E B}\atop \{a,b\}, \tree}\ \sum'_{\{\si \}}  
\prod_{j=1}^{n} \chi_{j}(\si) \prod_{j \not \in V} 
M^{-(l_j^1 + l_j^2 + l_j^3 + l_j^4)/4} \prod_{j \in V}   
M^{-(l_j^1 + l_j^2 + l_j^3)/4}  \ .
\ee

We enlarge the bound by suppressing the constraints on 
the sum over ${\{\si \}}$:

\be | p_n | \le{c^n \over n!} M^{-3i-3}
\sum_{\bar {\cal B}, {\cal E B}\atop \{a,b\}, \tree}\ \sum_{\{\si \}}  
\prod_{j=1}^{n} \chi_{j}(\si) \prod_{j \not \in V} 
M ^{-(l_j^1 + l_j^2 + l_j^3 + l_j^4)/4} \prod_{j \in V}   
M ^{-(l_j^1 + l_j^2 + l_j^3)/4} \ . \label{boubou}
\ee

Now we need the following lemma:

\begin{lemma}
Suppose the four sectors $\si_1, \si_2, \si_3, \si_4$ have
depths $l_1, l_2, l_3 $ and $l_4$. Then for fixed $\si_4$
\be
\sum_{\si_1, \si_2, \si_3} \chi (\si_1, \si_2, \si_3, \si_4) 
M^{-(l_1+l_2+l_3 )/4} \le c.i\ .
\ee
\end{lemma}
\noindent{\it Proof}\ \ 
Let us say that $\si_{j}$ collapses with 
$\si_{k}$ in the $\pm$ direction, and let us write $\si_{j}\simeq_{\pm} 
\si_{k}$ if $|s_{\pm,j} -s_{\pm,k} |\le 1$.
The function $\chi$ ensures two collapses,
one in each direction, for pairs with minimal
values of the corresponding $s$ indices. 
So it ensures that $\si_{j} \simeq_{+} \si_{k}$
and $\si_{j'}\simeq_{-} \si_{k'}$ for some $j\ne k$ and $j'\ne k'$.
Now let us make three remarks:

a) Since $l= s_{+}+s_{-}-i +2 \ge 0$,
for a given sector summing over $s_{+}$ knowing $s_{-}$ or vice versa
can be done  at the cost of a constant, using only a fraction of the 
decay $M^{-l/4}$. 

b) When a pair $j,k$ collapses in any direction, one element of the pair,
say $k$ is not the fixed sector ($k\ne 4$). Using remark a, 
for fixed $\si_{j}$  we can sum
over $\si_{k}$ at the cost of a constant,
using only a fraction of the decay $M^{-l_k/4}$. 

c) If a sector $\si_m$ does not collapse with any other sector in any 
direction, we must have some sector $j$ which collapses in 
both directions with an other sector. This means that $s_{+,m} \ge s_{+,j}$
{\it and} $s_{-,m} \ge s_{-,j}$. But then $l_m \ge l_j$. If $m\ne 4$
we have therefore 
\be M^{-l_m/4} =M^{-(l_m-l_j )/4} M^{-l_j /4} 
\le M^{-[(s_{+,m}-s_{+,j})-(s_{-,m}-s_{-,j})]/4}\ ,
\ee
and we can sum over $\si_{m}$ knowing $\si_{j}$ again at the cost
of a constant using a fraction of the decay $M^{-l_m/4}$.

Putting these three remarks together, we obtain the Lemma. 
Indeed by remark c) the eventual sums over sectors which collapse
with no other ones cost only constants. The sums over sectors that 
under collapse relations are connected to the fixed sector $\si_{4}$
also cost nothing by remark b). Finally there can remain at most one
non trivial equivalence class under collapse which does not contain
the fixed sector $\si_{4}$. We pay a single factor $i$ to  
fix say some $\si_{j,+}$ within this class, and using
again Remarks a) and b) we can achieve all other sums in that class
paying only some constants.  \qed

Now to prove the Theorem we return to (\ref{boubou}). 

We can use the same strategy as in [DR1-2] to sum over sectors,
following the natural ordering from leaves to root of our tree $\tree$.
Every vertex is related to the root vertex by a single path in
the tree which starts by a well defined half-line hooked to that vertex
called the root half-line.
We will pay for the sector sum at that vertex keeping the sector of
that root half line fixed. This last sector will be fixed later when
the vertex bearing the other half of the tree line 
associated to the root half-line is considered.
For the root vertex (of a vacuum graph) there will be still one last
sector to fix.
By Lemma 5, each sum over the sectors of a regular vertex (not in 
$V$)
costs therefore only $c.i$. Now from the data in ${\cal EB}$
we know how the vertices in $V$
group into pairs $\bar v_b, v_{b}$ associated to the bipeds $b$,
and by the momentum conservation constraint $\chi_{b}$ we know that
$\si_{\bar v_{b}}^4 \simeq \si_{v_{b}}^4$
and  $l_{\bar v_{b}}^4 \simeq l_{v_{b}}^4$.
Moreover we can choose the root to be external for a {\it maximal}
biped $b_{0}$ of ${\cal B}$, so that
either $\bar v_{b}$ or $v_{b}$ is the root (if $b=b_{0}$), or 
$\si_{\bar v_{b}}^4$ or  $\si_{v_{b}}^4$ is a root half-line (because
the root, being outside, is either left or right of the two point subgraph).
Suppose the root half-line is $\si_{\bar v_{b}}^4$. Since 
$\si_{\bar v_{b}}^4 \simeq \si_{v_{b}}^4$, we can sum 
over the sectors for both $\bar v_{b}$ and $v_{b}$ at once using Lemma 5
and we have
\be  \sum_{ \si_{\bar v_{b}}^4 {\rm \ fixed}, \ 
\si_{\bar v_{b}}^4 \simeq \si_{v_{b}}^4
\atop \si_{\bar v_{b}}^1, \si_{\bar v_{b}}^2, \si_{\bar v_{b}}^3 ,
\si_{v_{b}}^1, \si_{v_{b}}^2 ,\si_{v_{b}}^3} 
\chi_{\bar v_{b}}(\si)\chi_{v_{b}}(\si)
M ^{-(1/4)(l_{\bar v_{b}}^1 + l_{\bar v_{b}}^2 + l_{\bar v_{b}}^3 + 
l_{v_{b}}^1 + l_{v_{b}}^2 + l_{v_{b}}^3)}
\le (c. i)^2\ .
\ee
We have finally to pay for summing over the last root sector.
When the root vertex is not in $V$, 
we have a last sum to perform over some
$\si_{j}^4$ but we can use the decay factor $M^{-l_j^4}$
in (\ref{boubou}) to pay for it. So this last sum costs
an additional factor $i$. 
But when the root is say $v_{b_{0}}$, there is no $M^{-l_{v_{b}}^4}$
decay in (\ref{boubou}) and this last sum therefore
costs not a factor $i$ but a factor $i^2$.

Hence we arrive at:

\be | p_n | \le  M^{-2i}  i^{n+2}
\sum_{ {\cal B}, {\cal E B}, \{a,b\}, \tree} {c^{n} \over n!}\ .
\label{arriv}
\ee

The sum  over spin indices $\{a,b\}$ trivially
costs at most $4^{n}$, so from now on let us work
with fixed $\{a,b\}$. But to bound the sum
over ${\cal B}, {\cal E B}$ and $\tree$, one has to exploit the fact that
there is a balance: roughly speaking for ${\cal B}$ small the sum over
${\cal B}, {\cal E B}$ does not cost much and
we have Cayley's theorem which states that the number of possible trees 
$\tree$ at order $n$ is $n^{n-2}$; for ${\cal B}$ large (many bipeds)
the choice of which vertices belong to which biped may be costly,
but once it is done, the compatible trees are much fewer.
This is captured in the following lemma:

\begin{lemma}  There exists some constant $c$ such that
\be \sum_{{\cal B}, {\cal E B}, \tree} {1 \over n!} \le 
c^{n}\ .
\ee
\label{combinato}
\end{lemma}   

\noindent{\it Proof}\ \ 
Recall that $c$ is our generic name for a constant.
Let us call $N_{b}= |{\cal B}|$ the total number of bipeds.
For each $b \in {\cal B}$ (including the box $G$) let us 
call $d_{b}$ the number of links in ${\cal B}$ whose down end is $b$ 
and $n_{b} \le d_{b}$ the number of bare vertices 
that belong to $b$ and no smaller biped (dots in Figure \ref{bipf}
with a down link ending at $b$). Remark that $\sum_{b}n_{b}=n$ since
each dot belongs to some element of ${\cal B}$ (since we included $G$
in ${\cal B}$). Moreover $N_{b} \le n/2$ and 
$\sum_{b}d_{b}= n +N_{b}-1 \le 2n $. Therefore paying 
$c^{n}$ we can fix $N_{b}$ and the numbers $n_{b}$ and $d_{b}$
for each $b$.

We perform inductively the counting over the cardinal of the set
$(\bar {\cal B}, {\cal E B}, \tree)$ 
starting from the leaves in Figure \ref{bipf}
towards the root. To choose the $n_{b}$ vertices in each biped we have
to pay $n!/\prod_{b} n_b !$. To build the tree $\tree$ we build 
its restriction to each
reduced element of ${\cal B}$, which contains $d_{b}$ vertices ($n_{b}$
ordinary four point vertices and  $d_{b}-n_{b}$ reduced two point vertices).

Since by Cayley's theorem the number of trees on $n$ vertices is $n^{n-2}$  
hence bounded by $c^{n}n!$, the number of possible choices for $\tree$
is bounded by  $\prod_{b} c^{d_{b}} d_{b} !$, hence by 
$c^{n}\prod_{b}d_{b} !$. 
Finally to choose ${\cal E B}$ we fix for each $b$ the two fields
$\psi_{b}$ and $\bar \psi_{b}$. Since as remarked above
they must be hooked to the $n_{b}$ vertices that belong to $b$ 
and no smaller biped, the number of choices for ${\cal E B}$
is bounded by $\prod_{b} (4n_{b})^{2}$ hence by $c^{n}$ since 
$\sum_{b}n_{b}=n$. Multiplying all these numbers
we obtain a bound, but here comes the subtle point: in this way we
have counted $\prod_{b}  (d_{b}- n_b)!$ times each configuration
${\cal B}, {\cal E B}, T$.
Indeed the tree in Figure \ref{bipf} is {\it unlabeled}.
A permutation group with $\prod_{b}  (d_{b}- n_b)!$
acts on it, permuting at each fork $b$ the $d_{b}- n_b$ 
maximal bipeds in $b$, and each permutation on one
element of ${\cal B}, {\cal E B}, \tree$
built in the way described above gives again the same element. 

Hence what we have bounded is $ \sum_{{\cal B}, {\cal E B}, \tree} 
\prod_{b}  (d_{b}- n_b)!$ and we have obtained:
\be
\sum_{{\cal B}, {\cal E B}, \tree} \prod_{b}  (d_{b}- n_b)! 
\le c^{n} {n! \over \prod_{b} n_b !}
\prod_{b} d_{b} ! \le c^{n} n! \prod_{b}  (d_{b}- n_b)! 
\ee
which completes the proof of the Lemma. \qed

Combining this lemma and (\ref{arriv}), our final bound is
\be | p_n | \le i^{2} M^{-3i-3} (c.i)^n  \ .
\ee
This achieves the proof of Theorem \ref{radoneslice}. 
Remark that for a more general Schwinger function the prefactor
$i^2. M^{-3i-3}$ would be different but this has no influence
on the radius of convergence \qed

\section{Convergent contributions in the multislice theory}

To analyze the multislice theory, we remark first that  by Lemma 1,
integration over a vertex using the decay of a line with indices 
$i$ and $l$ costs $M^{2i+l}$. Therefore it is convenient to select the
multislice tree for a graph by optimizing over the index $r=I(i+l/2)$,
where $I$ is the integer part, so that $r$ remains integer.
From now on, we may forget the integer part $I$ which is inessential. 

In other words since $i$ in this section is no longer fixed,
we define the sectors as triplets $\si= (i, s_{+}, s_{-})$, 
with $1\le i \le i_{max}(T)$,
$0\le s_{+}\le i $, $0\le s_{-}\le i $, and $s_{+} + s_{-}\ge i-2 $.
The depth $l$ of a sector is still $l=s_{+} + s_{-}-i+2 $,
and the momentum cutoff $u_{si}$ for a sector is
\bqa u_{\si}(k_{0},k_+,k_{-})&=& 
u_i [ k_0^2+  4\cos^{2}(\pi k_{+}/2) \cos^{2} (\pi k_{-}/2) ] 
\nonumber\\
&&v_{s_{+}} [ \cos^{2} (\pi k_{+} /2 ) ]\;
v_{s_{-}} [ \cos^{2}(\pi k_{-}/2) ]\ .
\eqa
The propagator in sector $\si$ is in momentum space
\be C_{\si} (k_{0},k_+,k_{-}) = {u_{\si}(k_{0},k_+,k_{-})
\over ik_{0} - 2\cos(\pi k_{+}/2) \cos (\pi k_{-}/2) } \ .
\ee
The full propagator (with u.v. cutoff corresponding to the 
inessential removal of the slice $i=0$) is 
\be C=\sum_{\si}C_{\si}
=\sum_{r=1}^{r_{\max}(T)} C_{r} ,
\quad r_{\max}(T)=1+ 3i_{max}(T)/2 \ ,
\ee
the $r$-th slice of the propagator being defined as the sum
over all sectors with $i+l/2=r$:
\be
C_{r} = \sum_{\si \atop i(\si)+l(\si)/2 =r} C_{\si} = \sum_{l} C_{r,l}\ ,
\  C_{r,l}=\sum_{{\si \atop  i(\si)+l(\si)/2 =r}
\atop  l(\si) = l} C_{\si}\ .
\ee

Since we exclude in this section the graphs with two point subgraphs,
and any vacuum graph has such subgraphs, we shall formulate this time our
theorem for Schwinger functions $S_{2p}$ with $p\ge2$.
We perform the same Taylor jungle expansion as in the previous section,
but with respect to increasing values of the $r$ index, and obtain
(omitting the inessential dependence on the external momenta or positions): 
\be
S_{2p} = \sum_{n}  S_{2p,n}\; \la^n \ , \nonumber
\ee
\be
S_{2p,n} = {1\over n!}\sum_{\{a,b,\si\}}
\sum_{{\cal J}} \ep ({\cal J}) \prod_{v} \int dx_{v} 
\prod_{\ell\in \tree } \int_{0}^{1} dw_{\ell}
C_{i,\si_{\ell}} (x_{\ell}, y_{\ell})  
\det\nolimits_{left} (C_{i,\si}(w))  \ ,
\ee
where ${\cal J} = (\fr_0 \subset \fr_1 \subset ... \subset 
\fr_{r_{max}(T)} = \tree )$.

Knowing all the scales and sectors of all the fields, the connected 
components $G^{k}_{r}$ at each level $r$ again form a Gallavotti-Nicol{\`o}
tree for the inclusion relations. As seen below, in the 
$r$ space power counting is standard, namely the bipeds are
linearly divergent, and the ``quadrupeds'', namely the
non-trivial\footnote{Non-trivial here means ``not reduced to 
a single vertex''.}
components $G^{k}_{r}$ in the Gallavotti-Nicol{\`o} tree which
have $e(G^{k}_{r})=4$, are marginal. The bipeds require renormalization
and will be treated in another paper.
In this section we state two theorems: one for the 
``completely convergent'' part of the expansion, that is the one which has 
neither bipeds nor quadrupeds, and the other for the ``biped-free''
part of the expansion which has no biped but can have quadrupeds.
Indeed the first theorem is easier, so that order of presentation
seems more pedagogical.

Therefore we define the structure of all divergent
components as ${\cal B} \cup {\cal Q}$
with inclusion relations exactly as in Figure 
\ref{bipf}. ${\cal B}$ is the set of bipeds, hence
of connected components $G^{k}_{r}$ with $e(G^{k}_{r})=2$ and 
${\cal Q}$ is the set of quadrupeds, including the full graph $G$
pictured as the box in Figure \ref{bipf}, which may or may not
have four external legs, depending whether $p=2$ or $p>2$.
As in the previous section we also define ${\cal E Q}$ as the data
for the external legs of every quadruped $q \in {\cal Q}$.

We organize our sum as in the previous section and get the analog
of (\ref{form})
\be
S_{2p,n} = {1\over n!}\sum_{{{\cal B},  {\cal E B} \atop 
{\cal Q} , {\cal E Q} }\atop \{a,b\}, \tree}
\sum_{\{\si \}}' \ep (\tree) \prod_{v} \int dx_{v} 
 \prod_{\ell\in \tree} \int_{0}^{1} dw_{\ell}
C_{i,\si_{\ell}} (x_{\ell}, y_{\ell}) 
\det\nolimits_{left} (C_{i.\si}(w))  \ .\label{gene}
\ee
The completely convergent part of the functions 
$S_{2p}$, called $S_{2p}^{c} = \sum_{n} S_{2p,n}^{c} \lambda^{n}$, is now 
the sum over all contributions for which ${\cal B}={\cal Q}= \emptyset$,
namely $e(G^{k}_{r})>4 \ , \ \forall r,k$ (this requires
$p\ge 3$):
\be
S_{2p,n}^{c} = {1\over n!}\sum_{{\cal B} = {\cal Q}= \emptyset \atop 
\{a,b\} \tree }
\sum_{\{\si \}}' \ep (\tree) \prod_{v} \int dx_{v} 
 \prod_{\ell\in \tree} \int_{0}^{1} dw_{\ell}
C_{i,\si_{\ell}} (x_{\ell}, y_{\ell}) 
\det\nolimits_{left} (C_{i,\si}(w))  \ .
\ee

We can now state our second result\footnote{This 
definition of $S_{2p}^{c}$  has the disadvantage
not to be cutoff independent. Indeed it includes 
not only the sum of all the graphs without two and four point subgraphs,
which is a cutoff-independent object, 
but also some part of the amplitudes of graphs which do have
such two or four point subgraphs, namely those parts
in which these divergent subgraphs do not appear as connected
components $G_r^k$. However this definition
is the most natural one in the context of a multiscale expansion, and 
is similar to those of [FMRT],[DR1],[DMR].}:

\begin{theor} The functions $S_{2p}^{c}$ 
are analytic in $\la$ for $|\la \log T| \le c$ hence their radius of
convergence $R_T$ at temperature $T$ satisfies
\be  R_T  \ge c / | \log T| \ .
\ee
\label{radmulslice}
\end{theor}

\noindent{\it Proof} 
\ \ Before Gram's bound we now introduce only the momentum
conservation constraints at each bare vertex $j=1,...,n$:

\bqa
S_{2p,n}^{c} &= & {1\over n!}\sum_{{\cal B} = {\cal Q}= \emptyset \atop 
\{a,b\} \tree}
\sum_{\{\si \}}' \ep (\tree) \prod_{j=1}^{n}\chi_{j}(\{\si\}) 
\nonumber \\
&&\prod_{v} \int dx_{v} \prod_{\ell\in {\tree}} \int_{0}^{1} dw_{\ell}
C_{i,\si_{\ell}} (x_{\ell}, y_{\ell}) 
\det\nolimits_{left} (C_{i,\si}(w))  |_{x_{0}=0}\ .
\eqa

We can now apply 
Gram's inequality to the determinant as in the previous section, 
and integrate again over all positions of the vertices (save a few,
corresponding to the external arguments) using the decay of the tree 
propagators.

Since $(i+l)/2= r/2+l/4$, we obtain, exactly in the same way
as (\ref{absol1}), the bound
(holding one internal vertex fixed):
\be | S_{2p,n}^{c} | \le   {c^n \over n!} 
 \sum_{{\cal B} = {\cal Q}= \emptyset \atop 
\{a,b\} \tree} \sum_{\{\si \}}'
\prod_{j=1}^{n} \chi_{j}(\{\si\}) 
\prod_{\ell \in {\tree}} M^{2r_{\ell}}
\prod_{f} M^{-r_{f}/2-l_{f}/4}\ ,
\ee
where the product over $f$ runs again over all the fields
and antifields of the theory, and the sum over $\{\si \}$ is again
constrained to be compatible with the data $({\cal B} = {\cal Q}= \emptyset,
\{a,b\}, \tree )$, as indicated by the prime notation.

Using again (\ref{induc1}) and (\ref{induc2}) we obtain in a 
similar way:

\be  \prod_{f} M^{-r_{f}/2} = \prod_{r=0}^{r_{max}(T)}\prod_{k} 
M^{-e(G^{k}_{r}/2}\ ,
\ee
\be \prod_{\ell \in {\tree}} M^{2r_{\ell}} = M^{-2r_{max}(T) -2} 
\prod_{r=0}^{r_{max}(T)}\prod_{k} M^{2}\ ,
\ee
so that apart from a certain $n$ independent factor that cannot influence
the radius of convergence we get:

\be| S_{2p,n}^{c} | \le  {c^n \over n!} 
\sum_{{\cal B} = {\cal Q}= \emptyset\atop \{a,b\}, \tree} \sum_{\{\si \}}'
\prod_{j=1}^{n} \biggl[\chi_{j}(\{\si\})
M^{-(l_{j}^{1}+l_{j}^{2}+l_{j}^{3}+l_{j}^{4})/4} \biggr]
\prod_{r=0}^{r_{max}(T)} \prod_{k} M^{2-e(G^{k}_{r})/2}
\label{pow}
\ee
where at a given vertex $j$ we call $l_{j}^{1},..., l_{j}^{4}$
the depths of the four sectors hooked to $j$.

Since for all $r,k$ such that $G^{k}_{r}$ is non-trivial, 
$e(G^{k}_{r} \ge 6$, we have 
$2-e(G^{k}_{r})/2 \le - e(G^{k}_{r})/6$ for all such $r,k$. 
By a standard argument (see e.g. [R]) 
we conclude that if $r_{j}^{1}\le ... \le r_{j}^{4}$
are the $r$ scales of the four sectors hooked to $j$:
\be \prod_{r=0}^{r_{max}(T)} \prod_{k} M^{2-e(G^{k}_{r})/2}
\le \prod_{j} M^{-[(r_{j}^{2} -r_{j}^{1})+
(r_{j}^{3} -r_{j}^{1})+(r_{j}^{4} -r_{j}^{1})]/6}\ .
\ee
Therefore, forgetting the constraints on $\{\sigma \}$:

\be| S_{2p,n}^{c} | \le  {c^n \over n!} 
\sum_{\{a,b\}, \tree} \sum_{\{\si \}}
\prod_{j=1}^{n} \chi_{j}(\{\si\})
M^{- \sum_{k=1}^{4} l_{j}^{k} /4  -
\sum _{k\ne k'} | r_{j}^{k} -r_{j}^{k'} | /18 }\ .
\label{pogla}
\ee

Now with a fraction (say half) of the decay factor 
$\prod_{j} M^{-\sum _{k\ne k'} | r_{j}^{k} -r_{j}^{k'} | /18 }$
it is easy to perform the sum over all $r$ indices for all the fields
(just follow the tree like in the previous section: at each vertex three
$r$ indices can be summed holding the fourth fixed, which is the one of
the tree line going towards the root, and iterate until the root).
It is also possible with a fraction of the decay factor
$M^{- \sum_{k=1}^{4} l_{j}^{k} /4 }$ to sum over the indices $i$ once
the indices $r$ have been summed, since $i =r-l/2$. From now on
we consider therefore the former $i$ indices as fixed, although no longer
all equal as in the previous section). 

To sum over the sectors $s_{j}^{\pm}$ once scale indices $r$ and $i$
are fixed,
we have to be careful that case B of Lemma \ref{consmom} must now be used
since we are in a multislice case. The new possibility of case
B  of Lemma \ref{consmom} is that at a given vertex $j$ we can have 
$s_{j,\pm}^{1}=i_{j}^{1} < s_{j,\pm}^{k} $, 
$k=2,3,4$. Let us say that in this case the vertex $j$ 
is $\pm\;${\it special}.
In that case, since $s_{j,\pm}^{k} \le i_{j}^{k} \le r_{j}^{k}$, we have
\be |s_{j,\pm}^{k} - s_{j,\pm}^{1}| 
\le  r_{j}^{k} - i_{j}^{1} = r_{j}^{k} - r_{j}^{1} + l_{j}^{1}/2
\le |r_{j}^{k} - r_{j}^{1}| + l_{j}^{1}/2\ .
\ee
It is therefore easy to bound a fraction of the decay factor in (\ref{pogla})
by the product over the special vertices and directions of another decay
factor suitable for the summation of $s$ indices. For instance:
\be
\prod_{j\ \pm \; special} M^{- \sum_{k=1}^{4} l_{j}^{k} /8  -
\sum _{k\ne k'} | r_{j}^{k} -r_{j}^{k'} | /36 } \le \prod_{j \ \pm \; special}
 M^{- \sum _{k\ne k'} | s_{k,\pm} - s_{k',\pm} | /108 }\ .
\ee
Using this decay factor it is trivial to sum up all the $s_{\pm}$
indices of a {\it special} vertex, holding one fixed, namely the one
of the tree line going towards the root. The indices in the other
direction of the special vertex are easily summed with an other fraction
of the $l$ decay factor, namely $M^{- \sum_{k=1}^{4} l_{j}^{k} /8}$
Finally the sum over indices of the regular vertices which are special
neither in the plus nor in the minus direction can be handled
exactly as in the previous section, using up the remaining 
$\prod_{j \ {\rm not} 
\ special}  M^{- \sum_{k=1}^{4} l_{j}^{k} /8} $ factor.
Indeed their momentum conservation
is identical to Case A of Lemma 5. The corresponding
sums cost therefore at most $| c i_{max}(T) |^{n}$, 
hence at most $| c \log T |^{n}$

This achieves the proof of Theorem \ref{radmulslice}.
\qed

\medskip

Returning to (\ref{gene}), we define the biped-free part of the functions 
$S_{2p}$, called $S_{2p}^{bf} = \sum_{n} S_{2p,n}^{bf} \lambda^{n}$, as
the sum over all contributions for which ${\cal B} = \emptyset$
namely $e(G^{k}_{r})>2 \ , \ \forall r,k$ (this requires
$p\ge 2$):
\be
S_{2p,n}^{bf} = {1\over n!}\sum_{{\cal B} = \emptyset \atop 
{\cal Q}, {\cal EQ}, \{a,b\} \tree }
\sum_{\{\si \}}' \ep (\tree)   \prod_{v} \int dx_{v} 
\prod_{\ell\in \tree} \int_{0}^{1} dw_{\ell}
C_{i,\si_{\ell}} (x_{\ell}, y_{\ell})
\det\nolimits_{left} (C(w))  \ .
\ee

We can now state our third result\footnote{
This result involves again a cutoff-dependent quantity,
$S_{2p}^{bf}$, but with a little additional care it should be possible
to prove it also for a cutoff independent quantity, namely the
sum of all skeleton graphs. Indeed we know that extracting the self energy
part of the theory can be done constructively, at the cost
of a slightly more complicated expansion 
than a simple tree expansion (see [DR2], Appendix B).}:

\begin{theor} The functions $S_{2p}^{bf}$ 
are analytic in $\la$ for $|\la \log^{2} T| \le c$ hence their radius of
convergence $R_T$ at temperature $T$ satisfies
\be  R_T  \ge c / | \log^{2} T| \ .
\ee
\label{radbfslice}
\end{theor}

\noindent{\it Proof}\ \ 
Before Gram's bound we now introduce not only the momentum
conservation constraints at each bare vertex $j=1,...,n$ but also for each
quadruped $q \in {\cal Q}$:

\bqa
S_{2p,n}^{bf} &= & {1\over n!}\sum_{{\cal Q},{ cal EQ} \atop \{a,b\}, \tree}
\sum_{\{\si \}}' \ep (\tree) \prod_{j=1}^{n}\chi_{j}(\{\si\}) 
\prod_{q}\chi_{q}(\{\si\}) \nonumber \\
&& \prod_{v} \int dx_{v} \prod_{\ell\in {\tree}} \int_{0}^{1} dw_{\ell}
C_{i,\si_{\ell}} (x_{\ell}, y_{\ell}) 
\det\nolimits_{left} (C(w))  |_{x_{0}=0}\ .
\eqa

We can now apply Gram's inequality to the determinant as in the 
previous section, and integrate again over all positions of the 
vertices save one using the decay of the tree propagators. 
We obtain the analog of
(\ref{pow}):

\bqa | S_{2p,n}^{bf} | & \le &  {c^n \over n!} 
 \sum_{ {\cal Q} , {\cal E Q} \atop \{a,b\}, \tree} \sum_{\{\si \}}'
\prod_{q\in {\cal Q}} \chi_{q}(\{\si\})
\nonumber \\
&& 
\prod_{j=1}^{n} \chi_{j}(\{\si\})
M^{-(l_{j}^{1}+l_{j}^{2}+l_{j}^{3}+l_{j}^{4})/4} 
\prod_{r=0}^{r_{max}(T)} \prod_{k} M^{2-e(G^{k}_{r})/2}\ .
\label{powbf}
\eqa

We have no longer complete exponential decay between the scales of
the legs of any vertex. But the only missing piece corresponds
to the quadrupeds, for which in (\ref{powbf}) the factor $2-e(G^{k}_{r})/2$
is zero. This suggests an inductive bound which works inside
each reduced component $q/{\cal Q}$. The necessary data 
to perform this analysis are given in  $({\cal Q} , {\cal E Q})$.
Like in the previous section, let us introduce $n_{q}$ and $d_{q}$
as the number of ordinary vertices and the total number of vertices
in the reduced component $q/{\cal Q}$, so that $\sum_{q\in {\cal Q}}n_{q}=n$ 
and $\sum_{q\in {\cal Q}}(d_{q}-n_{q})=|{\cal Q}| \le n-1$
\footnote{The fact that any forest of quadrupeds 
has at most $n-1$ elements is a rather obvious statement, 
proved for instance in [CR, Lemma C1]).}.
Let us fix the largest scale $r_{q}$ inside $q$. Because 
the momentum conservation constraints for the external legs
of $q$ are included in (\ref{powbf}), the sums over $r$ scales
inside every reduced component $q/{\cal Q}$
(including the last one $G$ corresponding to the box in Figure \ref{bipf})
can be performed exactly like in the previous paragraph, at a cost of 
$c^{d_{q}}$ using the line with
scale $r_{q}$ as root for the $r$ indices summation. Similarly
the sums over the $s_{\pm}$ internal indices could be easily bounded
by $c^{d_{q}} |\log T|^{d_{q}}$,  using 
any given sector of an external line of $q$ 
as a root for these summations. But this bound is not optimal.
Let us prove that we can do better and perform these sums at a cost of
only  $c^{d_{q}} |\log T|^{d_{q}-1}$, holding all four external sectors
of the quadruped fixed. By remark c) in Lemma 5 and the analysis above, 
we pay a $|\log T|$ factor only for the vertices with two 
disjoint collapsing pairs. If one internal vertex of $q$ or
the external legs of $q$ do not have
disjoint collapsing pairs, we gain directly the necessary $|\log T|$ 
factor for $q$ later in the analysis. Otherwise, 
following the tree towards the root of the quadruped, like in Section 3,
we pay only at most $|\log T|^{d_{q}-1}$, because there is at least one 
sector sum fixed by the external data {\it in addition} to the root:
it is the one corresponding to the collapsing pair of
the external legs of $q$ {\it not} containing the root
\footnote{We remark that disjoint collapsing pairs
at a vertex correspond exactly to the combinatoric of a 
$\phi^{4}$ vector model. We know that for a Feynman graph we would pay
in fact $|\log T|^{cc}$ where $cc$ is the number of closed cycles.
It is well known that this number for a quadruped
with $d$ vertices is at most $d-1$. Our argument is a slight
adaptation of this fact, necessary because
we know only a spanning tree,
not the exact loop structure and closed cycles of a quadruped.}.
This proves the improved bound 
$c^{d_{q}} |\log T|^{d_{q}-1}$ for the $s_{\pm}$ internal summations.

Now for each $q$ we have also in addition to pay
a single additional $|\log T|$ factor to fix the scale $r_{q}$.
Multiplying all these factors,
we get for fixed $({\cal Q} , {\cal E Q}, \tree)$:
\bqa 
\sum_{\{\si \}}' &&
\prod_{q\in {\cal Q}} \chi_{q}(\{\si\})
\prod_{j=1}^{n} \chi_{j}(\{\si\})
M^{-(l_{j}^{1}+l_{j}^{2}+l_{j}^{3}+l_{j}^{4})/4} 
\prod_{r=0}^{r_{max}(T)} \prod_{k} M^{2-e(G^{k}_{r})/2} 
\nonumber\\
&&\le \prod_{q}
c^{d_{q}} |\log T|^{d_{q}} \le c^{n} 
|\log T |^{2n-1}\ .
\eqa
This completes the proof of Theorem \ref{radbfslice}, 
modulo the analog of Lemma \ref{combinato}:

\begin{lemma}  There exists some constant $c$ such that
\be \sum_{{\cal Q}, {\cal E Q}\atop\{a,b\} \tree} {1 \over n!} \le 
c^{n}\ .
\ee
\label{comb}
\end{lemma}   

\noindent{\it Proof}
The proof is identical to the one of Lemma \ref{combinato},
except for the little change that a given leg can now be external
to {\it several} quadrupeds. It is easy to take care of this detail:
in the sum over ${\cal EQ}$ there is simply a factor $d_{q}^{4}$
instead of $n_{b}^{2}$. But since $\sum_{q}d_{q}\le 2n$, it is again bounded
by $c^{n}$.\qed

\medskip

We expect the radius of analyticity for the full theory (with bipeds)
at temperature $T$ to satisfy the same bound as Theorem \ref{radbfslice}.
Indeed thanks to particle-hole symmetry, at half-filling  
the square Fermi-surface is preserved
under the RG flow. In contrast with the jellium case, there is therefore
no need to include any counterterm to formulate the analyticity theorem
for the full theory with bipeds.
Nevertheless power counting must be improved, i.e.
one has to transfer some internal convergence to the external legs,
hence to prove that by some Ward identity, the apparently divergent 
two point contributions are really convergent. This is postponed 
to a future paper.

\medskip
\noindent{\bf Acknowledgements}
\medskip

We thank F. Bonetto for many discussions and common work
on a preliminary version of this paper. 
In particular he found the elegant definition
of sectors in subsection II.2. We also thank C. Kopper for a critical
reading of the manuscript.

\medskip
\noindent{\large{\bf References}}
\medskip
\vskip.1cm

\noindent [AR1] A. Abdesselam and V. Rivasseau, Explicit Fermionic Cluster
Expansion, Lett. Math. Phys. {\bf 44} (1998) 77-88.
\vskip.1cm

\noindent [AR2] A. Abdesselam and V.  Rivasseau, Trees, forests and jungles: a
botanical garden for cluster expansions, in Constructive Physics, ed by
V. Rivasseau, Lecture Notes in Physics 446, Springer Verlag, 1995.
\vskip.1cm

\noindent [BG] G. Benfatto and G. Gallavotti,
Perturbation theory of the Fermi surface in a quantum liquid.
A general quasi particle formalism and one dimensional systems, 
 Journ. Stat. Phys. {\bf 59} (1990) 541.
\vskip.1cm

\noindent [BGPS] G.Benfatto,  G.Gallavotti, A.Procacci, B.Scoppola,
Commun. Math. Phys. {\bf 160}, 93 (1994).
\vskip.1cm

\noindent [BM] F.Bonetto, V.Mastropietro, Commun. Math. Phys. 
{\bf 172}, 57 (1995).
\vskip.1cm

\noindent [CR] C. de Calan and V. Rivasseau, Local Existence of the Borel
Transform in Euclidean $\Phi^{4}_{4}$,
Commun. Math. Phys. {\bf 82}, 69 (1981).
\vskip.1cm

\noindent [DMR] M. Disertori, J. Magnen and V. Rivasseau,  
Interacting Fermi liquid in three dimensions at finite temperature: 
Part I: Convergent Contributions, Annales Henri Poincar{\'e}, {\bf 2} (2001).
\vskip.1cm

\noindent [DR1] M. Disertori and V. Rivasseau, Interacting Fermi liquid 
in two dimensions at finite temperature, Part I: Convergent Attributions. 
Comm. Math. Phys. 215, 251, (2000)
\vskip.1cm

\noindent [DR2] M. Disertori and V. Rivasseau, Interacting Fermi liquid 
in two dimensions at finite temperature, Part II: Renormalization, Comm. 
Math. Phys. 215, 291 (2000)
\vskip.1cm

\noindent [DR3] M. Disertori and V. Rivasseau,
Continuous Constructive Fermionic Renormalization, 
Annales Henri Poincar{\'e}, {\bf 1}, 1 (2000). 
\vskip.1cm

\noindent [FKLT] J. Feldman, H. Kn{\"o}rrer, D. Lehmann and E. Trubowitz, 
Fermi Liquids in Two Space Time Dimensions, in {\it
  Constructive Physics} ed. by V. Rivasseau, Springer Lectures Notes in
Physics, Vol 446, 1995.
\vskip.1cm

\noindent [FMRT] J. Feldman, J. Magnen, V. Rivasseau and E. Trubowitz,
An infinite Volume Expansion for Many Fermion Green's Functions,
Helv. Phys. Acta {\bf 65}
(1992) 679.
\vskip.1cm

\noindent [FST] J. Feldman, M. Salmhofer and E. Trubowitz, Perturbation Theory 
around Non-nested Fermi Surfaces II.  Regularity of the Moving Fermi 
Surface, RPA Contributions, 
Comm. Pure. Appl. Math. {\bf 51} (1998) 1133;
Regularity of the Moving Fermi Surface, The Full Selfenergy,
to appear in Comm. Pure. Appl. Math. 
\vskip.1cm

\noindent [FT1]  J. Feldman and E. Trubowitz, 
Perturbation theory for Many Fermion Systems, Helv. Phys. Acta {\bf 63}
(1991) 156.
\vskip.1cm

\noindent [FT2]  J. Feldman and E. Trubowitz, The flow of an Electron-Phonon
System to the Superconducting State, Helv. Phys. Acta {\bf 64}
(1991) 213.
\vskip.1cm

\noindent [FMRT] J. Feldman, J. Magnen, V. Rivasseau and E. Trubowitz,
An infinite Volume Expansion for Many Fermion Green's Functions,
Helv. Phys. Acta {\bf 65}
(1992) 679.
\vskip.1cm

\noindent [G] M. Gevrey,  Sur la nature analytique des solutions des
 {\'e}quations aux d{\'e}riv{\'e}es partielles, (Ann. Scient. Ec. Norm. Sup., 3 s{\'e}rie.
t. 35, p. 129-190) in {\it Oeuvres de Maurice Gevrey}
 pp 243 , ed. CNRS (1970).
\vskip.1cm

\noindent [L] A. Lesniewski, Effective Action for the Yukawa$_{2}$ 
Quantum Field Theory, Commun. Math. Phys. {\bf 108}, 437 (1987).
\vskip.1cm

\noindent [MR] J. Magnen and V. Rivasseau, A Single 
Scale Infinite Volume Expansion for
Three Dimensional Many Fermion Green's Functions,
Math.  Phys. Electronic  Journal, Volume 1,  1995.
\vskip.1cm


\noindent [R] V. Rivasseau, 
From perturbative to constructive renormalization, 
Princeton University Press (1991).
\vskip.1cm

\noindent [S1] M. Salmhofer,
Continuous renormalization for Fermions and Fermi liquid theory,
Commun. Math. Phys. {\bf 194}, 249 (1998).

\noindent [S2] M. Salmhofer, Improved Power Counting and Fermi Surface
Renormalization, Rev. Math. Phys. {\bf 10}, 553 (1998).
\vskip.1cm


\end{document}